\documentclass{aa}  
\usepackage{graphicx}
\usepackage{txfonts}
\usepackage{mathrsfs}
\begin{document}

\title{A twelve-image gravitational lens system in the $z \simeq 0.84$
cluster Cl~J0152.7-1357 \thanks{Based on observations carried out with \emph{ESO VLT} (programs 69.A-0683, 72.A-0759, and 78.A-0746) and \emph{ESO NTT} (program 61.A-0676).}}

\author{C. Grillo \inst{1,2}, M. Lombardi \inst{1,2}, P. Rosati
  \inst{1}, G. Bertin \inst{2}, R. Gobat \inst{1}, R. Demarco \inst{3}, C. Lidman \inst{4}, V. Motta \inst{5}, and M.~Nonino \inst{6}}

\offprints{C. Grillo}

\institute{European Southern Observatory, Karl-Schwarzschild-Str. 2, D-85748, Garching bei M\"unchen, Germany\\
  \email{cgrillo@eso.org}
  \and
  Universit\`a degli Studi di Milano, Department of Physics,
  via Celoria 16, I-20133 Milan, Italy\\
  \and
  Department of Physics and Astronomy, Johns Hopkins University, Baltimore, MD 21218\\
  \and
European Southern Observatory, Alonso de Cordova 3107, Vitacura, Casilla 19011, Santiago 19, Chile\\
  \and
  Universidad de Valpara\'iso, Departamento de F\'isica y Astronomia, Avda. Gran Breta\~na 1111, Valpara\'iso, Chile\\
  \and
  INAF-Osservatorio Astronomico di Trieste, via G.B. Tiepolo 11, I-34141 Trieste, Italy
}

\authorrunning{C.~Grillo et al.}

\titlerunning{A twelve-image gravitational lens system in the $z
\simeq 0.84$ cluster Cl~J0152.7-1357}

\date{Received X X, X; accepted Y Y, Y}

\abstract
{Gravitational lens modeling is presented for the first discovered
example of a three-component source for which each component is
quadruply imaged. The lens is a massive galaxy member of the
cluster Cl J0152.7-1357 at $z\simeq 0.84$ .}
{Taking advantage of this exceptional configuration and of the
excellent angular resolution of the \emph{HST Advanced Camera for Surveys} (\emph{ACS}), we measure the
properties of the lens. In particular, the lensing mass estimates of the galaxy are compared to those from stellar dynamics and multiwavelength photometry.}
{Several parametric macroscopic models were developed for the lens galaxy, starting from pointlike to extended image models. By combining lensing, stellar dynamics, photometry, and spectroscopy, we find an allowed range of values for the redshift of the source and the required minimum amount of dark matter enclosed within the disk defined by the Einstein ring of the lens.}
{For a lens model in terms of a singular isothermal sphere with
external shear, the Einstein radius is found to be $R\mathrm{_{E} =
9.54 \pm 0.15}$ kpc. The external shear points to the cluster's northern mass
peak. The unknown redshift of the source is determined to be higher than 1.9 and lower than 2.9. Our estimate of the lensing projected total mass inside the Einstein radius, $M_{\mathrm{len}}(R \le 9.54\mathrm{\,kpc})$, depends on the source distance and lies
between $4.6$ and $6.2 \times \mathrm{10^{11}\,M_{\sun}}$. This result turns out to be compatible with the dynamical estimate based on an isothermal model. By considering the constraint on the stellar mass-to-light ratio that comes from the evolution of the Fundamental Plane, we can exclude the possibility that at more than 4 $\sigma$ level the total mass enclosed inside the Einstein ring is only luminous matter. Moreover, the photometric-stellar mass measurement within the Einstein radius gives a minimum value of 50\% (1 $\sigma$) for the dark-to-total matter fraction.}
{The lensing analysis has allowed us to investigate the distribution of mass of the deflector, also providing some interesting indications on scales that are larger (cluster) and smaller (substructure) than the Einstein radius of the lens galaxy. The combination of different diagnostics has proved to be essential in determining quantities that otherwise would have not been directly measurable (with only the currently available data): the redshift of the source and the amount of dark matter in the lens.}


\maketitle
%

\section{Introduction}

Accurate measurements of the mass present in galaxies in the forms of dark and visible matter define the empirical framework for the theory of galaxy formation and evolution. The knowledge of the galaxy mass function is essential for testing galaxy formation models (e.g., Hernquist \& Springel \cite{her03}; De Lucia et al. \cite{del06}). Among the estimators of total mass in elliptical galaxies, stellar dynamics is particularly valuable in the local Universe (e.g., Saglia et al. \cite{sag92}; Gerhard et al. \cite{ger}), while gravitational lensing offers the best direct diagnostics at high redshift (e.g., Kochanek \cite{koc95}; Impey et al. \cite{imp98}). Recently, spectral energy distribution (SED) fitting methods have achieved a level of precision sufficiently high to give reliable estimates of the mass contained in the form of stars (e.g., Fontana et al. \cite{fon04}; Rocca-Volmerange et al. \cite{roc04}; Saracco et al. \cite{sar04}; Grillo et al. \cite{gri08b}). By combining (e.g., Trott \& Webster \cite{tro02}; Treu et al. \cite{tre06}; Koopmans et al. \cite{koo06}) or by comparing (e.g., Drory et al. \cite{dro04}; Ferreras et al. \cite{fer05}; Rettura et al. \cite{ret06}) different mass estimators, the internal structure of galaxies can be investigated. In particular, the total density distribution and the fraction of mass in the form of dark matter can be determined.

Gravitational lensing has proved to be a unique tool to measure the
(total) projected mass of galaxies and clusters of galaxies on radial
scales from kiloparsecs to megaparsecs (e.g., Kochanek \cite{koc95}; Broadhurst et
al. \cite{broad}), and in particular, to study in detail the relationship between dark and luminous
matter for these systems.

In this respect, strong lensing models of multiply imaged sources are
especially important because the large number of observational constraints can lead to robust and detailed mass models.  The most complex system known so far
is a ten-image radio gravitational lens (Sykes et al. \cite{sykes};
Nair \cite{nair}).

Cl J0152.7-1357 is a rich, irregular, and X-ray luminous distant ($z
\simeq 0.84$) cluster (see Della Ceca et al. \cite{del00}; Demarco et al. \cite{dem05}). Its inner regions exhibit spectacular strong
gravitational lensing features in the form of multiple images and
arcs. Strong lensing of background galaxies is also detected on
smaller scales, around some individual cluster members. A new and
particularly interesting example is studied in this paper.

The paper is organized as follows. In Sect. 2, the observations of the galaxy cluster Cl J0152.7-1357 are described. In Sect. 3, we address the strong lensing analysis of the system presented in Fig.~\ref{Col}. We report the derived information on the redshift of the multiply imaged source in Sect. 4. Then, in Sect. 5, different stellar and total mass measurements of the lens galaxy are compared. Finally, in Sect. 6 we summarize the results obtained in this paper. 
Throughout this work we assume the following values for the cosmological
parameters: $H_{0}=70 \mbox{ km s}^{-1} \mbox{ Mpc}^{-1}$, $\Omega_{m}
= 0.3$, and $\Omega_{\Lambda} = 0.7$; in this model $1\arcsec$
corresponds to a linear size of $7.57\mathrm{\,kpc}$ at the lens
plane.

\section{Observations}

Cl J0152.7-1357 was discovered in the \emph{ROSAT} Deep Cluster Survey (RDCS; Rosati et al. \cite{ros98}; Della Ceca et al. \cite{del00}) as an extended source with a double core structure. Spectroscopy of six galaxies (Ebeling et al. \cite{ebe00}) confirmed the cluster and gave a redshift of $z \simeq 0.83$. The X-ray properties of Cl J0152.7-1357 have been studied in greater detail by \emph{BeppoSAX} (Della Ceca et al. \cite{del00}), \emph{XMM-Newton}, and \emph{Chandra} (Maughan et al. \cite{mau03}; Huo et al. \cite{huo}). The X-ray observations were used to derive the X-ray luminosity of the cluster and the temperature ($kT \approx 6$ keV, with typical errors of 20$\%$) and metallicity of the intra-cluster medium (ICM). The same data provided some evidence of a possible merger of the two main subclumps, one to the northeast (hereafter northern subcluster) and the other to the southwest (hereafter southern subcluster). These two structures are at a projected distance of $\approx$~1.6\arcmin $\,$ (corresponding to $\approx$ 730 kpc at the cluster redshift.)

The cluster was observed with the Wide Field Channel of the
\emph{ACS} in November and December 2002. It was imaged in
the F625W (\emph{r}), F775W (\emph{i}), and F850LP (\emph{z})
bandpasses as part of a Guaranteed Time Observation program (proposal
9290). The observations were done in a 2$\times$2 mosaic pattern
allowing for a $\approx 50 \arcsec$ overlap between pointings, with
integrated exposure of $\approx 4800 \mbox{ s}$ (see Jee et al. \cite{jee}; Blakeslee et al. \cite{bla}). By using \emph{ACS} observations, a weak lensing analysis provided a detailed mass map of the cluster (Jee et al. \cite{jee}). The mass reconstruction showed complicated substructure in spatial agreement with the two peaks of the ICM traced by the X-ray emission. Moreover, the X-ray ($2.4^{+0.4}_{-0.3}\times 10^{14}\,M_{\odot}$) and weak lensing [$(2.1 \pm 0.3)\times 10^{14}\,M_{\odot}$] estimates of the total mass enclosed within small radii ($\approx 50$\arcsec) turned out to be consistent.

Further multi-band optical and near-IR imaging observations of Cl J0152.7-1357 were performed to select targets for spectroscopy (see Demarco et al. \cite{dem05}; J\o rgensen et al. \cite{jorgensen}). Optical photometry in the $B$-, $V$-, $R$-, and $I$-bands was obtained with the Low Resolution Imaging Spectrometer (\emph{LRIS}) at the W. M. Keck Observatory. The \emph{LRIS} images cover a region of 4.9\arcmin $\times$ 6.5\arcmin. The cluster was observed in the $V$-, $R$-, and $I$~-bands with the FOcal Reducer and low dispersion Spectrograph (\emph{FORS1}) at the \emph{VLT}. The \emph{FORS1} images cover a region of 6.8\arcmin $\times$ 6.8\arcmin. Imaging was also done in the three $r$\arcmin-, $i$\arcmin-, and $z$\arcmin-bands with the Gemini Multi-Object Spectrograph on Gemini North (\emph{GMOS-N}). The \emph{GMOS-N} images cover a region of 5.5\arcmin $\times$ 5.5\arcmin. The cluster was imaged in the near-infrared ($J$- and $Ks$-bands) with \emph{SofI}, on the \emph{NTT} at the Cerro La Silla Observatory. The \emph{SofI} images cover a region of 4.9\arcmin $\times$ 4.9\arcmin.

Cl J0152.7-1357 was at the center of an extensive spectroscopic campaign carried out with both \emph{FORS1} and \emph{FORS2} at the \emph{VLT} (Demarco et al. \cite{dem05}). A total of 11 masks for multi-object spectroscopy were used, covering the wavelength range 4000-10000 $\AA$, with a total exposure time of $\approx$ 29.1 hr. More than 200 redshifts were measured and 102 galaxies were confirmed as cluster members. Spectroscopic observations in the wavelength range 5000-10000 $\AA$ were also performed with one mask of the \emph{GMOS-N}, with a total exposure time of $\approx$ 21.7 hr (J\o rgensen et al. \cite{jorgensen}). Redshifts and central velocity dispersions were measured for 41 galaxies, 29 of which were identified as cluster members. From these analyses it was found that the irregular distribution of the cluster members follows the extended X-ray emission. Furthermore, the mean redshift of the cluster was measured $z = 0.837 \pm 0.001$, and the values of velocity dispersion of $\approx$ 919 and $\approx$ 737 km s$^{-1}$ were estimated for the northern and southern subclusters, respectively.

In conclusion, X-ray, weak lensing, and dynamical studies all agree in finding in this system that light traces total mass, which is concentrated in two main peaks. The inner parts of these subclusters coincide with the two regions where strong gravitational lensing mostly occurs.

Finally, the temperature and the mass [$M(<65\arcsec) = (2.1\pm0.7)\times 10^{14}\,M_{\odot}$] of the cluster were also measured from Sunyaev-Zel'dovich effect observations, obtained with the interferometers of the Berkeley-Illinois-Maryland Association (\emph{BIMA}; Joy et al. \cite{joy01}). A good agreement with the results inferred from the X-ray analyses was found.

\begin{figure}
  \centering
  \includegraphics[height=3.15cm]{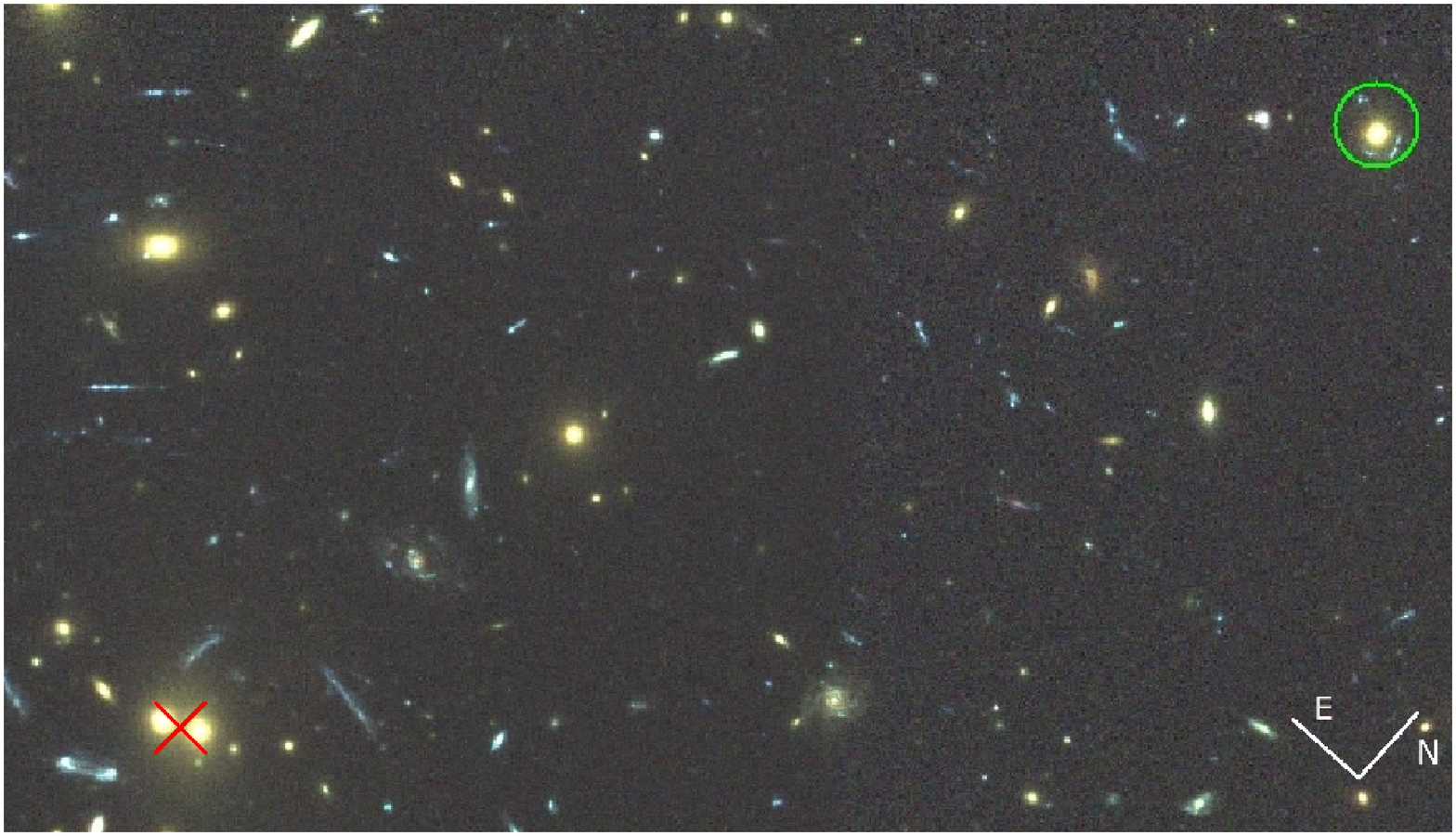}
  \includegraphics[width=3.15cm]{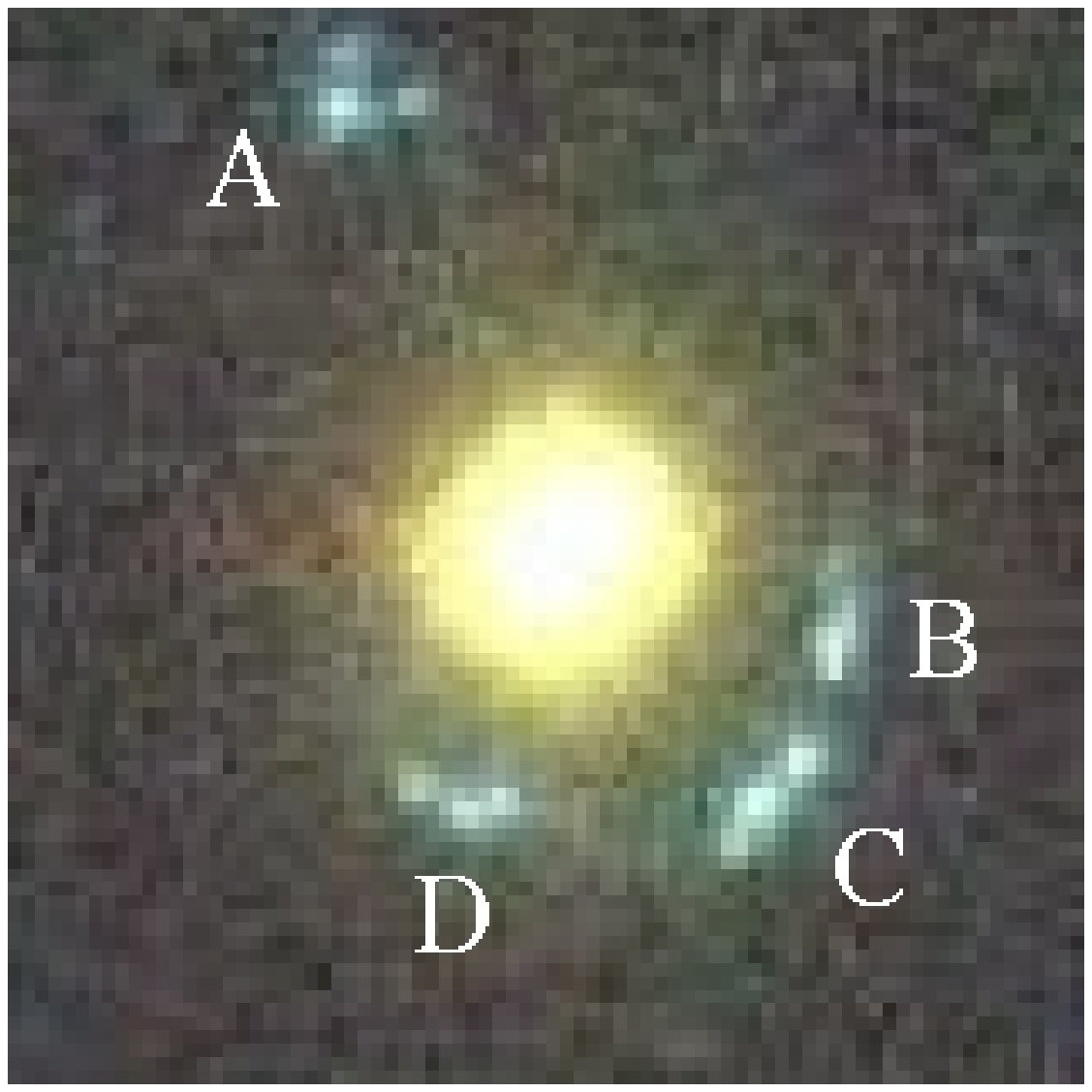}
  \caption{\emph{Left}: Color image of a small portion ($\sim1.1\arcmin$ $\times$ $\sim0.7\arcmin$) of the northern subcluster created using the
  \emph{riz} \emph{HST/ACS} filters. The green circle on the top 
  right, at a projected distance of 488 kpc and in direction $26.4\degr$ NE with respect to the centroid of the northern mass clump, which corresponds to two central galaxies (indicated by the red cross on the bottom left), shows the position of the strong lensing
  system considered in this paper. \emph{Right}: Zoom-in view of a
  $4\arcsec \times 4\arcsec$ field around the lens galaxy at
  $z_{l}=0.82$.} 
  \label{Col}
\end{figure}

In Fig.~\ref{Col} we show a color-composite image of a small region in the northern subcluster obtained by combining the three \emph{HST/ACS} filters. The strong lensing system considered in this work is indicated by a green circle, on the left, and shown in detail, on the right of Fig.~\ref{Col}. The cluster lens galaxy is located at a projected distance of 488 kpc from the centroid of the northern mass clump (which corresponds to two central galaxies), where the most multiple images and arcs produced by Cl J0152.7-1357 were detected. This geometrical configuration will justify the need for an external shear component in the lensing models of the next section. From its spectral and photometrical properties, the lens can be safely classified as a bright elliptical galaxy located at a redshift $z_{l}$ of $0.82$ and with a stellar central velocity dispersion $\sigma_{0}$ of $239\pm21$ km s$^{-1}$ (see J\o rgensen et al. \cite{jorgensen}). Despite the outstanding observational efforts that will be described in Sect. 4.2, the redshift of the quadruply-imaged source $z_{s}$ was not secured. It was only possible to give an upper limit: $z_{s} \leq 2.9$. The measured photometric properties of the lensing system are summarized in Tables \ref{table:1}, \ref{table:2}, and \ref{table:3}; more details on the properties of this system will be given below.



\begin{table}
\caption{The lens galaxy.}
\label{table:1}
\centering
\begin{tabular}{c c c c c c c}
\hline\hline
R.A. & Dec. & $z_{l}$ & $e$ & P.A. & $r_\mathrm{AB}$ &
$R_{\mathrm{e}}$ \\
(J2000) & (J2000) & & & (\degr) & (mag) & (\arcsec) \\
\hline
01:52:42.43 & $-$13:56:17.96 & 0.82 & 0.08 & 51.7 & 22.84 & 0.28 \\
\hline
\end{tabular}
\end{table}

\begin{table}
\caption{Photometry of the four images.}
\label{table:2}
\centering
\begin{tabular}{c c c c c}
\hline\hline Object & $x_{1}$ $^{\mathrm{a}}$ & $x_{2}$
$^{\mathrm{a}}$ & distance $^{\mathrm{a}}$ &
$r_\mathrm{AB}$ \\ & (\arcsec) & (\arcsec) & (\arcsec) & (mag) \\
\hline A & -0.76 & 1.69 & 1.86 & 25.58 \\ B & 1.04 & -0.28 & 1.07 & 25.99 \\ C & 0.79 & -0.89 & 1.19 & 25.32 \\ D &
-0.31 & -0.92 & 0.97 & 25.87 \\ \hline
\end{tabular}
\begin{list}{}{}
\item[$^{\mathrm{a}}$] With respect to the galaxy center.
\end{list}
\end{table}

\begin{table}
\caption{Astrometry of the sub-images.}
\label{table:3}
\centering
\begin{tabular}{c c c c c c c}
\hline\hline Object & $\mathrm{A}_{1}$ & $\mathrm{A}_{2}$ &
$\mathrm{A}_{3}$ & $\mathrm{B}_{1}$ &$\mathrm{B}_{2}$ &
$\mathrm{B}_{3}$ \\ \hline $x_{1}\,(\arcsec)$ $^{\mathrm{a}}$ & -0.81
& -0.80 & -0.58 & 1.04 & 1.04 & 1.04 \\ $x_{2}\,(\arcsec)$
$^{\mathrm{a}}$ & 1.62 & 1.82 & 1.67 & -0.28 & -0.28 & -0.28 \\
\noalign{\smallskip} \hline\hline Object & $\mathrm{C}_{1}$ &
$\mathrm{C}_{2}$ & $\mathrm{C}_{3}$ & $\mathrm{D}_{1}$ &
$\mathrm{D}_{2}$ & $\mathrm{D}_{3}$ \\ \hline $x_{1}\,(\arcsec)$
$^{\mathrm{a}}$ & 0.72 & 0.66 & 0.90 & -0.34 & -0.24 & -0.54 \\
$x_{2}\,(\arcsec)$ $^{\mathrm{a}}$ & -0.93 & -1.10 & -0.77 & -0.98 &
-0.93 & -0.88 \\ \hline
\end{tabular}
\begin{list}{}{}
\item[$^{\mathrm{a}}$] With respect to the galaxy center.
\end{list}
\end{table}

\section{Gravitational lensing models}
At first, we model the source as a single point-like object, which is
lensed approximately in the four images of Table~\ref{table:2}. Then,
we employ a three-component source, trying to reproduce the twelve
images of Table~\ref{table:3} (each of the previous four images, hence also the source, is considered here as a triple object). Finally, we study models with extended sources, in order to compare the model-predicted luminosity distribution of the images directly with the observations.

\subsection{Four images}
For the lens we consider three different models: a singular
isothermal sphere (SIS), a singular isothermal ellipsoid (SIE), and a
singular isothermal sphere with external shear (SIS+ES). An SIS is
characterized by three parameters: the Einstein angle
$\theta_\mathrm{E}$ and the two source coordinates
$(y_{1},y_{2})$. The remaining models involve two additional parameters: the
ellipticity $e=1-b/a$ and the position angle $\theta_{e}$ for an SIE;
the shear $\gamma$ and its position angle $\theta_{\gamma}$ for an
SIS+ES (see Keeton \cite{keet}). Initially, each model is analysed by fixing the lens center $(x_{1_{l}},x_{2_{l}})$ to the measured galaxy
position; in a second stage, the lens coordinates are also taken as free parameters. Varying these parameters and the position of the source, we minimize the chi-square function
\begin{equation}
\chi^{2} = \sum_{i=1}^{4} 
\frac{\|{\vec{x}^{i}_\mathrm{obs}-{\vec{x}^{i}}\|^{2}}}
     {\sigma_{x}^{2}} \, ,
\end{equation}
where $\vec{x}^{i}_\mathrm{obs}$ is the position vector of the $i$-th observed
image (see Table~\ref{table:2}), $\vec{x}^{i}$ is the corresponding position predicted by the
model, and $\sigma_{x}$ is the position uncertainty. This last
quantity is fixed to one image pixel (0.05$\arcsec$) because the
centroids of the extended images are not well defined.

\begin{table}
\caption{The best-fit parameters for the four-image models.}
\label{table:4}
\centering
\begin{tabular}{c c c c c c c c}
\hline\hline Model & $\theta_{\mathrm{E}}$ & $x_{1_{l}}$ & $x_{2_{l}}$
& $e/\gamma$ $^{\mathrm{a}}$ & $\theta_{e/\gamma}$ $^{\mathrm{a}}$ &
$\chi^{2} $ & dof $^{\mathrm{c}}$\\ & (\arcsec) & (\arcsec) &
(\arcsec) & & (\degr) & & \\ \hline SIE $^{\mathrm{b}}$ & 1.91
& & & 0.486 & 28.3 & 14.3 & 3 \\ SIE & 2.55 & 0.15 & -0.17 & 0.699 &
28.1 & 1.77 & 1 \\ SIS+ES $^{\mathrm{b}}$ & 1.22 & & & 0.198 & 29.1 &
6.70 & 3 \\ SIS+ES & 1.23 & 0.06 & 0.05 & 0.185 & 28.6 & 1.00 & 1 \\
\hline 
\end{tabular}
\begin{list}{}{}
\item[$^{\mathrm{a}}$] Ellipticity or external shear values, depending
on the model.
\item[$^{\mathrm{b}}$] The lens center is fixed to the galaxy center.
\item[$^{\mathrm{c}}$] Number of degrees of freedom.
\end{list}
\end{table}

\begin{figure}
\centering
\includegraphics[width=0.49\textwidth]{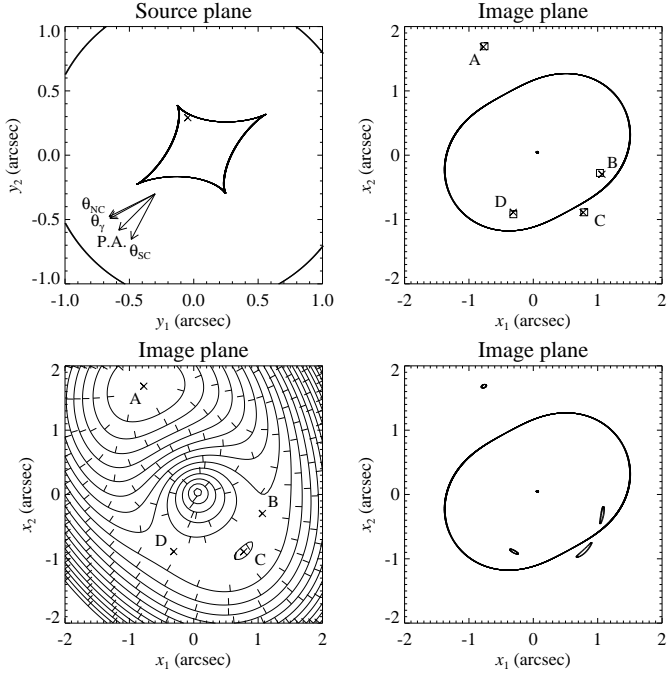}
\caption{Best four-image SIS+ES model. \emph{Top left}: Source plane
with caustics. The predicted source position is indicated by the cross. The arrows display the directions of the northern cluster clump ($\mathrm{\theta_{NC}}$), of the external shear predicted by the model ($\mathrm{\theta_{\gamma}}$), of the galaxy ellipticity position angle (P.A.), and of the southern cluster clump ($\mathrm{\theta_{SC}}$). \emph{Top right}: Image
plane with critical curves. The observed (squares) and predicted
(crosses) image positions are shown. \emph{Bottom left}: Contour levels of the
Fermat potential. The images are two
minima (A, C) and two saddle points (B, D). \emph{Bottom right}: Predicted deformation of the four images obtained by locating a small round source in the position illustrated on the top left panel.}
\label{Multi}%
\end{figure}

The best-fit parameters are shown in Table~\ref{table:4}. These models have different properties. The SIS is not able to reproduce the correct number of images, and thus the parameters of this model are not registered. The SIE, with the center coordinates as free
parameters, has a low $\chi^{2}$, but the high value of the lens ellipticity, compared to that measured for the galaxy (see Table \ref{table:1}),
suggests that this model is inadequate.  The least $\chi^{2}$ value
is achieved by an SIS+ES, setting the lens center as a free parameter;
the second best model is the SIS+ES with fixed center (the higher
$\chi^2$ is partially compensated for by the higher number of degrees of freedom, dof). 
In Fig.~\ref{Multi} we plot the source and image planes with caustics
and critical curves for the last model of Table \ref{table:4}. The positions of the four images, typical of a cusp configuration, are there compared with the positions predicted by the model. We also show the Fermat potential (see Schneider et al. \cite{schneider}), its stationary points, corresponding to the image locations (two minima and two saddle points with, respectively, the same and the opposite parity of the source), and the predicted level of distorsion of the images from a round source.

We notice that the high values found for the external shear are plausible in a massive cluster, as is the case of Cl J0152.7-1357. For instance, a lens cluster with an Eistein radius of 200 kpc causes an external shear of 0.2 at a distance of 500 kpc from the cluster center (a simplified SIS model is assumed here to model the lens). We will see in Sect. 5 that the values of the external shear determined in this section are fully consistent with the mass estimates of the northern subcluster measured through studies of X-ray emission (Huo et al. \cite{huo}), weak lensing (Jee et al. \cite{jee}), and cluster member dynamics (Demarco et al. \cite{dem05}). Moreover, it is interesting to remark that the position angle of the ellipticity for the SIE and that of the shear for the SIS+ES differ by about one degree. This common direction points towards the closest cluster mass peak, coinciding with the northern clump center of the cluster shown in Fig.~\ref{Col}.  

\subsection{Twelve images}
We start here from a triple source (labelled by an index running
from one to three) lensed into twelve images angularly close to the
measured positions of Table~\ref{table:3}. Considering the results of the previous subsection, a three-component source, for which each component is imaged four times, is the most natural assumption on the source structure. Three of the observed images,
$\mathrm{B}_{i}$, are so close to each other that they cannot be
deblended; hence, we just measure their bright center. The position
uncertainty of each sub-image is chosen to be one image pixel, as
before, except for $\mathrm{B}_{i}$ which is taken five times larger, by analyzing the peculiar luminosity distribution of the pixels. We
model the lens as an SIE and an SIS+ES.

\begin{table}
\caption{The best-fit parameters for the twelve-image models.}
\label{table:6}
\centering
\begin{tabular}{c c c c c c c c}
\hline\hline Model & $\theta_{\mathrm{E}}$ & $x_{1_{l}}$ & $x_{2_{l}}$
& $e/\gamma$ $^{\mathrm{a}}$ & $\theta_{e/\gamma}$ $^{\mathrm{a}}$ &
$\chi^{2} $ & dof $^{\mathrm{c}}$ \\ & (\arcsec) & (\arcsec) & (\arcsec) & & (\degr) &
& \\ \hline SIE $^{\mathrm{b}}$ & 1.87 & & & 0.441 & 25.1 & 37.5 & 15
\\ SIE & 2.19 & 0.10 & -0.10 & 0.588 & 26.6 & 23.9 & 13 \\ SIS+ES
$^{\mathrm{b}}$ & 1.26 & & & 0.179 & 26.1 & 23.3 & 15 \\ SIS+ES & 1.27
& 0.03 & 0.04 & 0.169 & 26.7 & 21.6 & 13 \\ \hline
\end{tabular}
\begin{list}{}{}
\item[$^{\mathrm{a}}$] Ellipticity or external shear values, depending on the model.
\item[$^{\mathrm{b}}$] The lens center is fixed to the galaxy center.
\item[$^{\mathrm{c}}$] Number of degrees of freedom.
\end{list}
\end{table}

\begin{figure}
\centering
\includegraphics[width=0.49\textwidth]{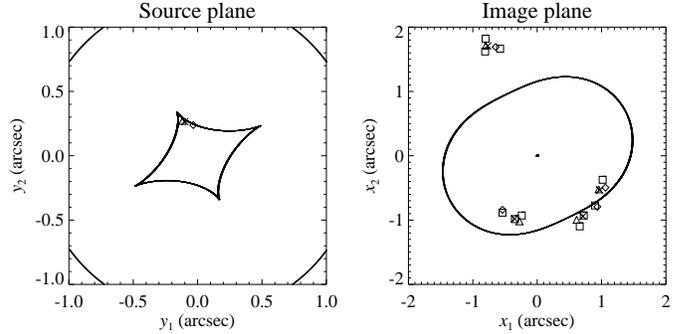}
\caption{Best twelve-image SIS+ES model. \emph{Left}: Source plane
with caustics. The predicted components of the source are
shown. \emph{Right}: Image plane with critical curves. The observed
(squares) and predicted (triangles, crosses, and diamonds) images are
shown.}
\label{Twelve}%
\end{figure}

The $\chi^{2}$ minimization results are summarized in Table \ref{table:6}. The best $\chi^{2}$, 23.3, is achieved by an SIS+ES, a model
consisting only of three parameters. If the lens center is also taken
as a free parameter, this reduces the $\chi^{2}$ by 1.7, but
lowers by two the number of dof. The source and image planes with caustics and
critical curves for the third model of Table \ref{table:6} are plotted
in Fig. \ref{Twelve}.

We note that the value of the Einstein angle of the SIS is very similar to that of the best four-image model and that the values of magnitude and orientation of the external shear give good evidence about the northern subcluster mass peak, as already mentioned in the last subsection.

Then, we determine the statistical uncertainties on the Einstein angle and on the external shear, starting from two hundred $\chi^{2}$ minimizations on simulated data sets. In particular, we extract image positions from Gaussian distributions centered on the measured
values and with standard deviations equal to the position
uncertainties reported previously, and for each set we search for the best-fit parameters.

\begin{figure}
\centering
\includegraphics[width=0.49\textwidth]{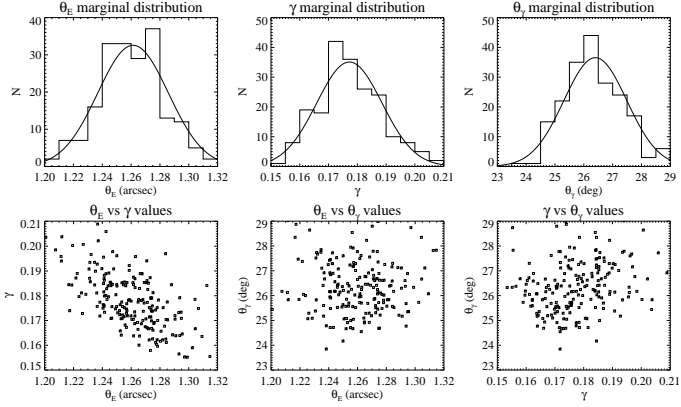}
\caption{Results of the $\chi^{2}$ minimizations of 200 Monte-Carlo
generated data sets. \emph{Top}: Marginal distributions of the model
parameters with their approximate normal distributions. \emph{Bottom}:
The joint density distributions show the parameter correlations.}
\label{Stat}
\end{figure}

The parameter distributions are shown in Fig. \ref{Stat} and the results are presented in Table \ref{table:5}. The marginal density functions are approximately Gaussian and the joint density functions reveal correlations (for definitions see Cowan \cite{cowan}): the correlation coefficients are $r_{\theta_{\mathrm{E}},\gamma}=-0.64$, $r_{\theta_{\mathrm{E}},\theta_{\gamma}}=-0.03$, and $r_{\gamma,\theta_{\gamma}}=0.19$. The origin of the significant anti-correlation of the value of the Einstein angle and the magnitude of the external shear will be explained in Sect.~5.1. We notice here that, at $z_{l}=0.82$, an Einstein angle of $(1.26\pm0.02)$~\arcsec $\,$corresponds to an Einstein radius ($R_{\mathrm{E}}$) of $(9.54 \pm 0.15)$~kpc. Moreover, we remark that the low uncertainty values on the parameters suggest that the measurements are accurate and robust.

The model just presented is valuable in describing the lens properties
with only three parameters, although it cannot reproduce the complex image configuration very accurately. In order to improve the
agreement between the observed and the reconstructed image geometry we have 
tried more refined models, without gaining any real improvement. For instance, some ellipticity is inserted in the SIS+ES model, but, as expected from the known degeneracy between external shear and ellipticity (see Witt et
al. \cite{witt}), we cannot find a lower value of the $\chi^{2}$.

\begin{table}
\caption{The best-fit parameters with the relative marginalized errors for the
twelve-image SIS+ES model.}
\label{table:5}
\centering
\begin{tabular}{c c c}
\hline\hline $\theta_{\mathrm{E}}$ (\arcsec) & $\gamma$ &
$\theta_{\gamma}$ (\degr) \\ \hline
$1.26 \pm 0.02$ & $0.179 \pm 0.011$ &
$26.4 \pm 1.1$ \\ \hline
\end{tabular}
\end{table}

We interpret the external shear as due to the mass distribution on the cluster scale. So, instead of employing an external shear component, we decide to model both the northern subcluster (the most relevant mass clump for this study) and the galaxy as SISs. The description of the subcluster in terms of an SIS is only a first-order approximation, but it is adequate to represent the mass distribution at distances from the center large compared to the Einstein radius of the clump (below we will prove that this statement is valid in our system). The two Einstein angles are the only needed parameters (hence, the number of dof is here 16). In fact, the first center is set equal to the average position of the two brightest galaxies of the northern subcluster (see Fig. \ref{Col}), the second one to the center of the lens galaxy. We measure a $\chi^{2}$ of 25.5, and Einstein angles for the subcluster ($\theta_{\mathrm{E_{C}}}$) and for the galaxy ($\theta_{\mathrm{E_{G}}}$) of 19.6\arcsec $\,$($R_{\mathrm{E_{C}}}=$ 148 kpc) and 1.07\arcsec $\,$($R_{\mathrm{E_{G}}}=$ 8.10 kpc), respectively. This latter value will be used in the following as a conservative lower limit of the galaxy Einstein angle.

Finally, as far as the source properties are concerned, the predicted three-component source has a linear extent of about $0.09\arcsec$, equivalent to approximately 0.7 kpc at a hypothetical redshifts of 2.5. In addition, the total absolute magnification of the sub-images ($\mu = |\mu_{\mathrm{A}}| + |\mu_{\mathrm{B}}| + |\mu_{\mathrm{C}}| + |\mu_{\mathrm{D}}|$, where $\mu_{i}$ is the magnification of the $i$-th image, with a positive or negative sign if the image is, respectively, a minimum or a saddle point of the Fermat potential) is 27.9, a reasonable value in the vicinity of critical lines.

Taking into consideration the irregular and compact shape and the blue color of the source, it is very likely that this object is a high-redshift star-forming region.


\subsection{Extended images}

As a final step, we employ an extended parametric approach. For this purpose, we have developed an \emph{ad hoc} algorithm, which reconstructs the properties of both a lens and a source by comparing the observed and the predicted luminous intensity of an array of pixels on the image plane. Before presenting the results, we describe this technique shortly, since it is not as well-known as the point-like parametric (Keeton \cite{kee01}) or the non-parametric (Saha \& Williams \cite{sah97}; Koopmans \cite{koo05}) approaches.

In principle, the method is straightforward, because it exploits only the ray-tracing equation, $\vec{y}(\vec{x}\,)=\vec{x}-\vec{\alpha}(\vec{x}\,)$, and the surface brightness conservation in lensing, $I(\vec{x}\,)=I^{s}\big(\vec{y}(\vec{x}\,)\big)$ (for more details see Schneider et al. \cite{schneider}). First, the image and the source planes are divided into sub-pixels. In particular, each CCD pixel of the image plane is splitted into 16 sub-pixels. Furthermore, in order to deal with large magnified images, the pixel size on the source plane is fixed to 1/256 the pixel size of the CCD. Then, starting values for the parameters of the models that describe the lens ($\vec{p}_{l}\,$) and the luminosity distribution of the source ($\vec{p}_{s}\,$) are assigned. According to the tentative lens model, each point of the image grid is ray-traced into the source plane and is associated with the closest point of the source grid. Finally, the images of the lensed source are obtained by giving to the image sub-pixels the values of the source sub-pixels matched in the previous step. The effect of the point spread function (PSF), $I(\vec{x}\,)=\big(\mathrm{PSF}*I^{s}\big)\big(\vec{y}(\vec{x}\,)\big)$, and the effect of the sub-pixels (anti-aliasing) are taken into account. The best parameters for the adopted lens and source models are found by minimizing the following chi-square function
\begin{equation}
\widetilde\chi^{2}(\vec{p}_{l},\vec{p}_{s}) = \frac{
\sum_{N_{\mathrm{pix}}}\frac{\big(I_{\mathrm{obs}}(\vec{x}\,)-I(\vec{x}\,)\big)^2}{\sigma_{I}^{2}}}{N_{\mathrm{pix}}} \, ,
\end{equation}
where $N_{\mathrm{pix}}$ is the total number of pixels of the observation considered in the modeling, $I_{\mathrm{obs}}(\vec{x}\,)$ and $I(\vec{x}\,)$ are, respectively, the intensity observed and predicted by the model in the CCD pixel located at $\vec{x}$, and $\sigma_{I}$ is the standard deviation of the intensity evaluated on a blank field of the CCD (i.e., the noise). The method has been tested on artificial but plausible lens systems and has provided good results for the values of the reconstructed parameters. 

Following the best models discussed above, we represent the lens as an SIS+ES and, on first approximation, we model the source as the sum of three gaussian functions with independent parameters.

\begin{table}
\caption{The best-fit parameters for the extended-image model.}
\label{table:8}
\centering
\begin{tabular}{c c c c c}
\hline\hline Model & $\theta_{\mathrm{E}}$ & $\gamma$ & $\theta_{\gamma}$ & $\widetilde\chi^{2}$ \\ & ($\arcsec$) & & ($\degr$) & \\ \hline SIS+ES $^{\mathrm{a}}$ & 1.25 & 0.190 & 28.6 & 1.35 \\ \hline
\end{tabular}
\begin{list}{}{}
\item[$^{\mathrm{a}}$] The lens center is fixed to the galaxy center.
\end{list}
\end{table}
\begin{figure}
\centering
\includegraphics[width=0.24\textwidth]{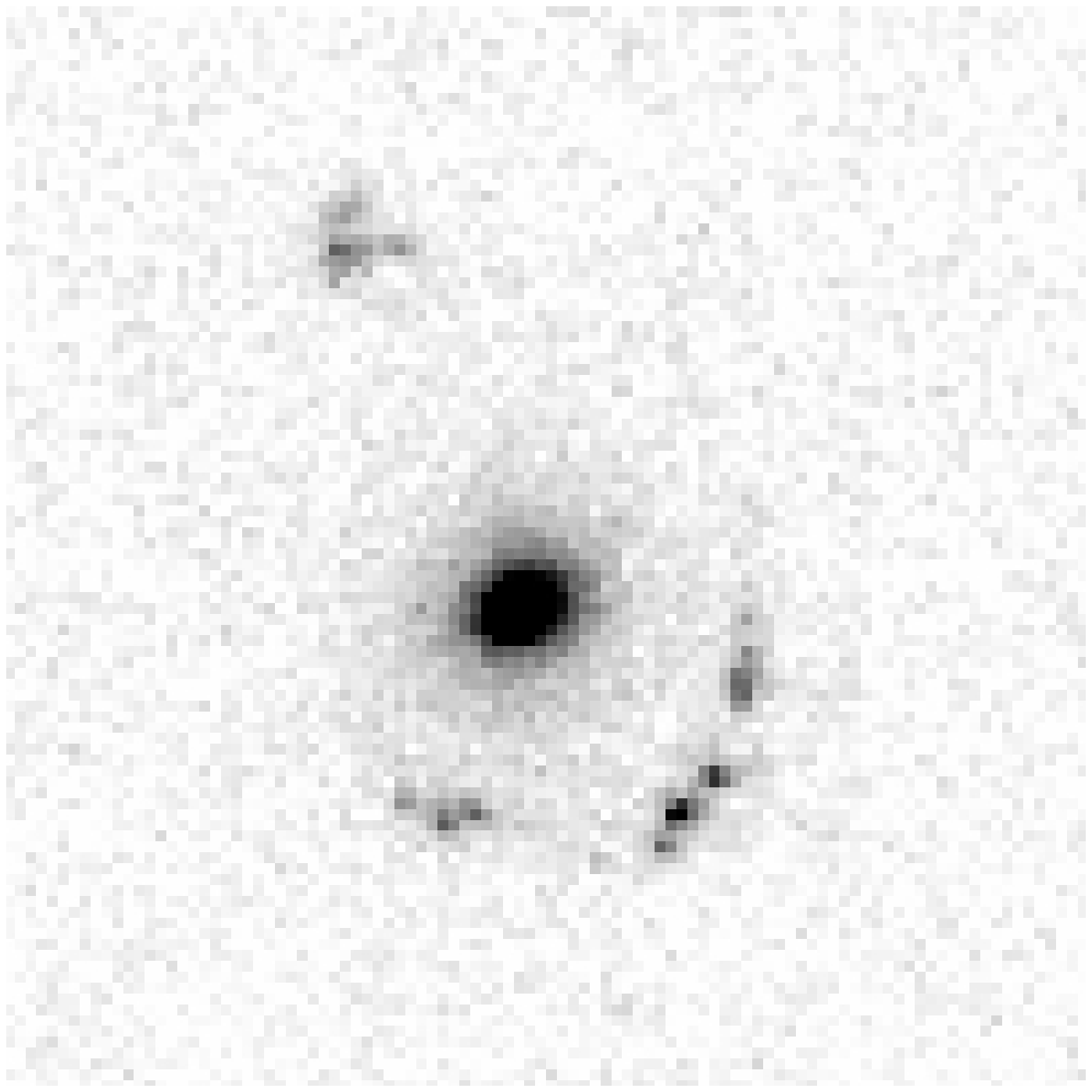}
\includegraphics[width=0.24\textwidth]{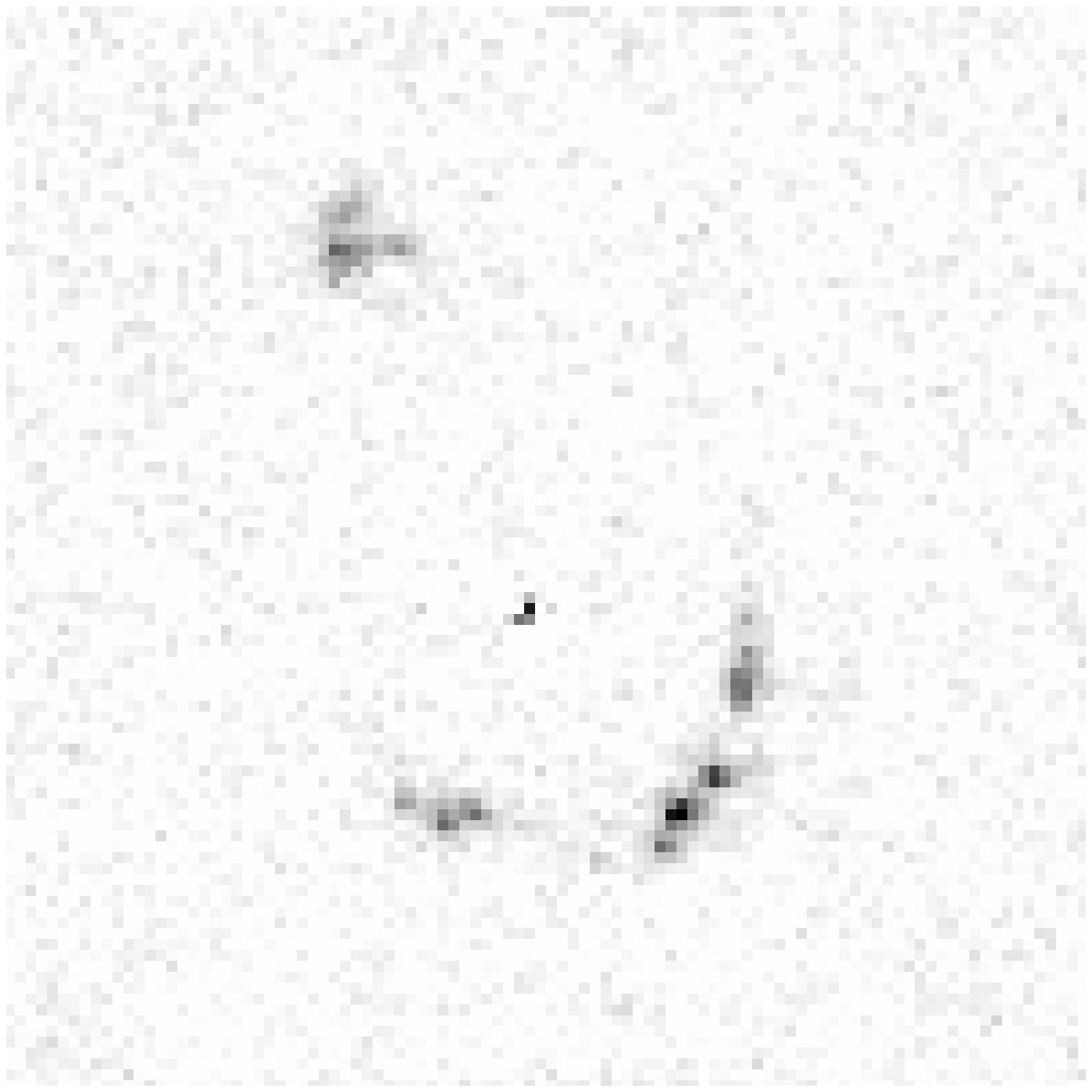}
\includegraphics[width=0.24\textwidth]{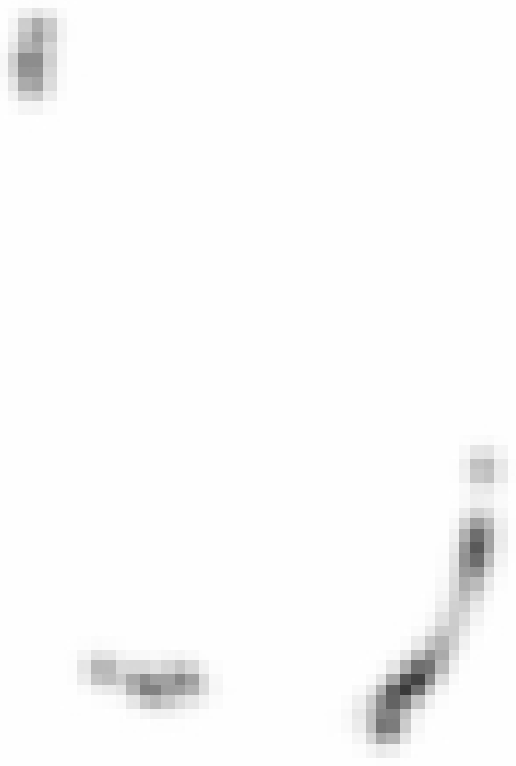}
\includegraphics[width=0.24\textwidth]{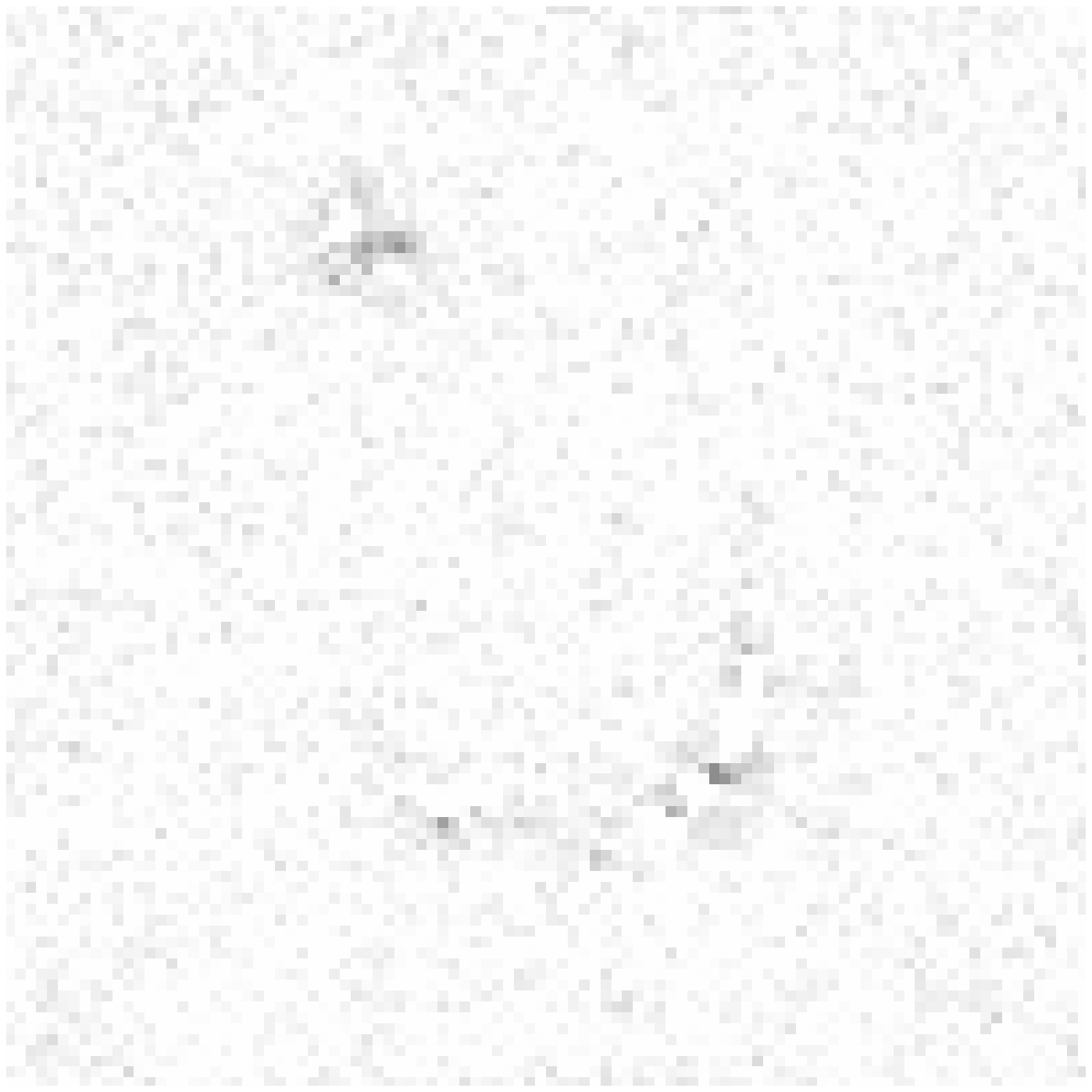}
\caption{Best extended-image SIS+ES model. \emph{Top left:} The observed \emph{HST/ACS} \emph{r} field (5\arcsec across). \emph{Top right:} The same image after the subtraction of an elliptical model fitted on the luminosity profile of the lens galaxy. Some residuals of the galaxy luminosity are still visible in the inner region. \emph{Bottom left:} The best reconstruction of the lensed system. \emph{Bottom right:} The residuals after the subtraction of the model predicted (third panel) from the observed (second panel) images. The central residuals due to the lens galaxy subtraction have been masked here.} 
\label{Ext}
\end{figure}
The best-fit parameters and model resulting from the application of our algorithm are shown in Table \ref{table:8} and Fig. \ref{Ext}. The properties of the lens are essentially the same as those obtained previously; on the other hand, we note slight variations in the values of the source parameters.
In fact, the values of the Einstein angle and of the external shear are consistent with those of the best point-like model (see Table \ref{table:5}). The largest separation among the centers of the three gaussians now gives a predicted linear extent for the source of about $0.18\arcsec$, which corresponds to approximately 1.4 kpc at a redshift of 2.5. The total absolute magnification, i.e. the integrated intensity of the images divided by that of the source, is on the order of 14.9. Table \ref{table:9} summarizes the signed (positive or negative if the nature of the two Fermat stationary points of each pair is, respectively, the same or not) flux ratios of the images A ($f_{\mathrm{A}}/f_{\mathrm{C}}$), B ($f_{\mathrm{B}}/f_{\mathrm{C}}$), and D ($f_{\mathrm{D}}/f_{\mathrm{C}}$), with respect to the most magnified image C, and the cusp ratio ($R_{\mathrm{cusp}}$) for the closest triplet of images B, C, and D. This last quantity, which is introduced by Keeton et al. \cite{kee03}, can be written as
\begin{equation}
(R_{\mathrm{cusp}})_{\mathrm{BCD}} = \frac{\Big|1+\frac{f_{\mathrm{B}}}{f_{\mathrm{C}}}+\frac{f_{\mathrm{D}}}{f_{\mathrm{C}}}\Big|}
{1+\Big|\frac{f_{\mathrm{B}}}{f_{\mathrm{C}}}\Big|+\Big|\frac{f_{\mathrm{D}}}{f_{\mathrm{C}}}\Big|} \, .
\end{equation}
In Table \ref{table:9} we report the observed values and their uncertainties, assuming a 10$\%$ error on each flux, and the model-predicted values of the previously defined quantities. For the three images B, C, and D, we measure an opening angle $\theta$ of  $93.6\degr$ and a dimensionless separation $d/\theta_{\mathrm{E}}$ of $1.20$ (for definitions see Keeton et al. \cite{kee03}). 
\begin{table}
\caption{Signed flux and cusp ratios.}
\label{table:9}
\centering
\begin{tabular}{c c c c c}
\hline\hline  & $f_{\mathrm{A}}/f_{\mathrm{C}}$ & $f_{\mathrm{B}}/f_{\mathrm{C}}$ & $f_{\mathrm{D}}/f_{\mathrm{C}}$ & $(R_{\mathrm{cusp}})_{\mathrm{BCD}}$ \\ \hline 
Obs. & $0.79\pm0.11$ & $-0.54\pm0.08$ & $-0.60\pm0.08$ & $0.07\pm0.06$\\
Mod. & 0.41 & $-0.59$ & $-0.56$ & 0.07 \\ \hline
\end{tabular}
\end{table}

The image positions taken into consideration in the pointlike models are supposed to coincide approximately with the centers of the images of the extended models. Nonetheless, the former analyses consider essentially only the regions where the images are maximally amplified by lensing, whereas the latter studies probe also the smaller amplification regime. Therefore, we do not have any obvious explanation for the different predicted sizes of the source; while, the difference between the values of the total magnification found here and in the previous paragraph can be easily understood.

For an ideal cusp catastrophe, the cusp ratio should exactly vanish (Schneider et al. \cite{schneider}), but, since real lenses are not ideal cusps, this general property is expected to hold only approximately, depending on the opening angle and the dimensionless separation of the triplet considered (for a detailed study see Keeton et al. \cite{kee03}). Surprisingly, most of the observed lens systems show a significant violation of the cusp relation (known also as flux ratio ``anomaly''; e.g., see Kent \& Falco \cite{ken88}; Sluse et al. \cite{slu03}). This fact can be explained by assuming the presence of small-scale structure in the lens galaxy, on a scale comparable to the separation between the images (Mao \& Schneider \cite{mao98}). Although we measure a cusp ratio which is consistent with zero (and thus, given the measured values of the opening angle and of the dimensionless separation, the cusp relation is here satisfied), a possible clue on the existence of small perturbations in the lens potential might be the significant inconsistency between the observed and the predicted flux ratio for image A (see Table \ref{table:9} and the last panel of Fig. \ref{Ext}).

\section{The source redshift}

A precise value for the redshift of the lensed source is not available
yet. Its measurement is very difficult because of the faintness of the
object and its proximity to the bright lens galaxy (see Table
\ref{table:2}). However, some information about this redshift can be
extracted from different diagnostics, as explained below.

\subsection{Gravitational lensing and stellar dynamics combined}

The Einstein angle for an SIS is explicitly given by
\begin{equation}
\theta_{\mathrm{E}}=4\pi\,\bigg(\frac{\sigma_{\mathrm{SIS}}}{c}\bigg)^{2}\frac{D_{ls}}{D_{os}}\,,
\label{eq:4}
\end{equation}
where $\sigma_{\mathrm{SIS}}$ is the lens ``velocity dispersion'', $c$ is the
speed of light, $D_{ls}$ and $D_{os}$ are the lens-source and the
observer-source angular diameter distances, respectively. It follows from Eq. (\ref{eq:4}) that $\theta_{\mathrm{E}}$ may be
interpreted as a function of both the lens mass $M$ ($M \propto
\sigma_{\mathrm{SIS}}^{2}r_{t}$) and the source redshift $z_{s}$ ($D_{ls}$,
$D_{os}$). A lower limit for this redshift can be obtained by combining gravitational lensing and stellar dynamics. 

\begin{figure}
\centering
\includegraphics[width=0.30\textwidth]{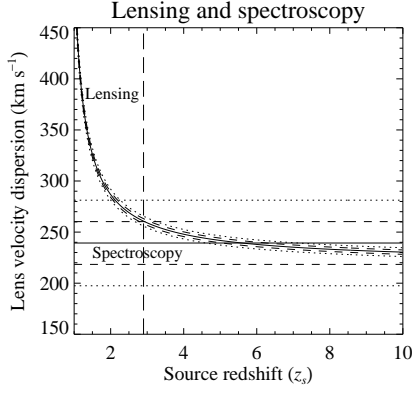}
\caption{The source redshift obtained from gravitational lensing and stellar dynamics.
The Einstein angle, from lensing analysis, and the velocity dispersion
of the galaxy, from spectroscopic measurements, combine to give a lower limit for the redshift of the source.
The dashed and dotted lines represent 1 and 2 $\sigma$ error bars, respectively. The vertical long-dashed line shows the spectroscopic upper limit of the source redshift discussed in Sect. 4.2}
\label{Red1}%
\end{figure}

In fact, several tests (Kochanek \cite{koc93}, \cite{koc94}; Treu et al. \cite{tre06}; Grillo et al. \cite{gri08}) have proved that in elliptical galaxies the stellar central velocity dispersion ($\sigma_{0}$) is a good estimate of the velocity dispersion of a one-component isothermal model ($\sigma_{\mathrm{SIS}}$). A spectroscopic measurement of $\sigma_{0}$ for the lens galaxy is presented by J\o rgensen et al. \cite{jorgensen} ($\sigma_{0}=239\pm21$ km s$^{-1}$). Starting from this result and from the value of the galaxy Einstein angle of the previous section, in Fig. \ref{Red1}, the velocity dispersions from gravitational lensing and from stellar dynamics are found to be compatible at a confidence level of $95\%$ if $z_{s}\ge 2.1$ and at $99\%$ if $z_{s}\ge 1.9$.

\subsection{Spectroscopic information}

\begin{figure}
\centering
\includegraphics[width=0.25\textwidth]{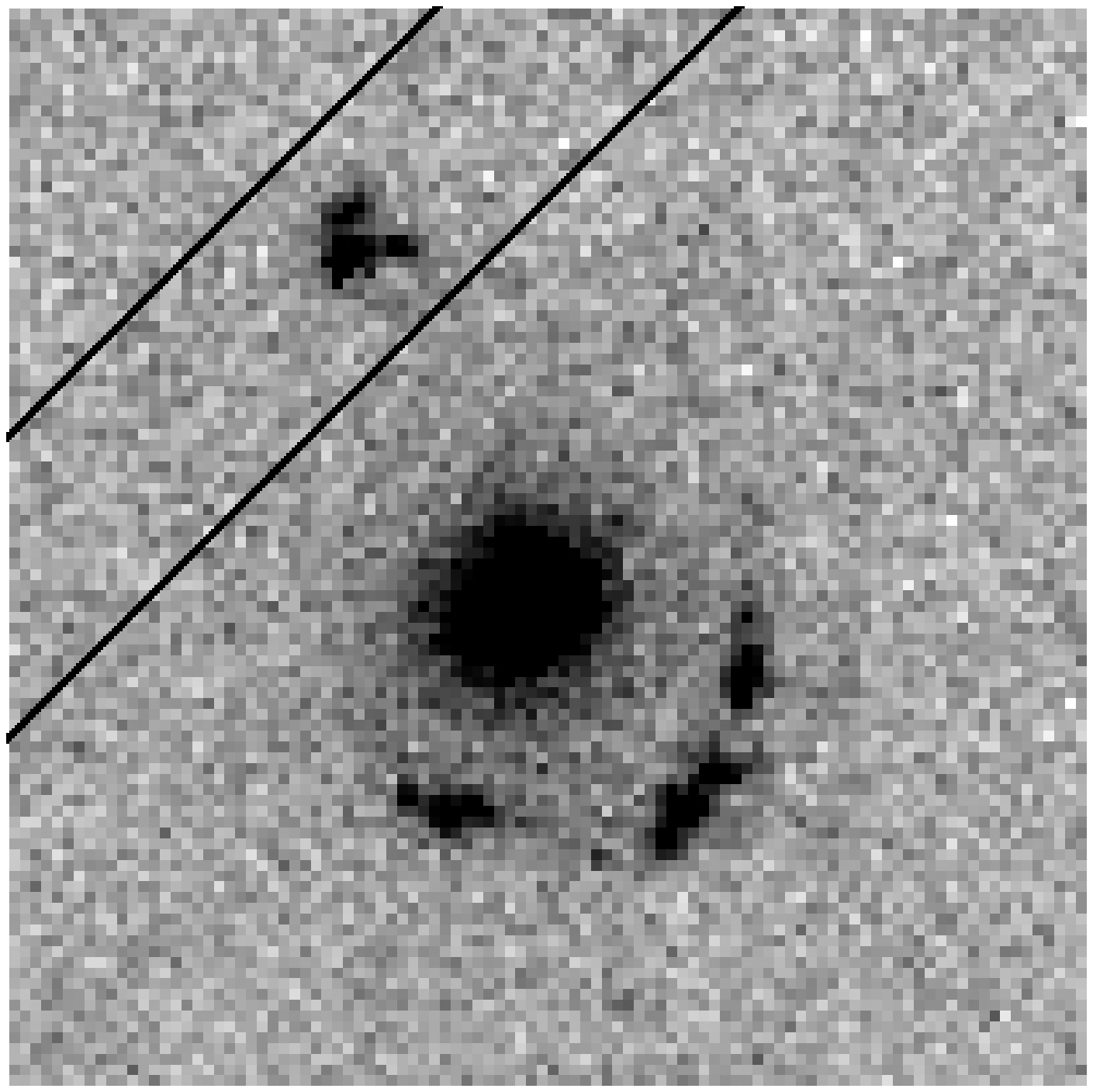}
\includegraphics[width=0.49\textwidth]{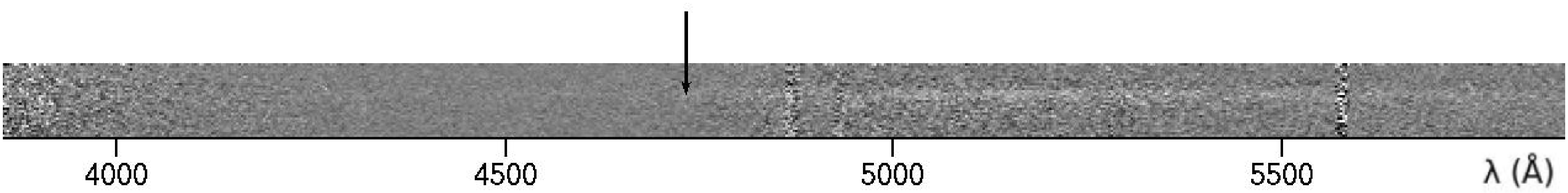}
\caption{Spectroscopic measurements of object A. \emph{Top}: Position of the 1\arcsec slit. \emph{Bottom}: Composite 2D sky-subtracted spectrum obtained with \emph{VLT/FORS2} in 5230 s. The arrow shows the Lyman alpha drop at 4730~$\AA$.}
\label{spe}
\end{figure} 

The spectra of objects A, B, and C were obtained with \emph{VLT/FORS2}, as part of an extensive observational campaign carried out on the galaxy cluster (see Demarco et al. \cite{dem05} for more details on the spectroscopic analysis of other objects in Cl J0152.7-1357). An unambiguous estimate of the source redshift was not possible from the spectra of images B and C, due to the contamination
from the overwhelming lens galaxy light in $0.8\arcsec$ seeing conditions.
The spectrum of image A was taken in three different exposures, with a total integration time of 5230 s, with the 300V grism, using the same mask geometry and the same slit characteristics. The position of the slit and the combined 2D sky-subtracted spectrum are shown in Fig. \ref{spe}. Although the
signal is low, no significant emission feature is observed, the
continuum is visible down to 4730 $\AA$, however the cross-correlation
with template spectra does not yield satisfactory results. If we interpret this limit as the Lyman alpha drop, blueward to the Lyman alpha emission line, then we can infer that the redshift of the source ($z_{s}$) is lower than 2.9.

\section{Mass measurements}

In this section we first determine the lens galaxy total (luminous+dark) mass, based on the previously discussed lensing models and on the published (J\o rgensen et al. \cite{jorgensen}) stellar velocity dispersion measurement. Next, we measure the lens galaxy stellar mass, based on multiwavelength photometry spectral template fitting. This allows us to infer a lower limit for the dark matter fraction enclosed inside the Einstein radius of the deflector.

\subsection{Lensing estimates}

\begin{figure}
\centering
\includegraphics[width=0.49\textwidth]{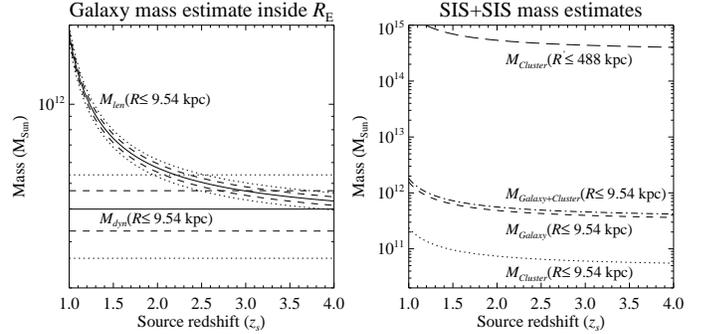}
\caption{The galaxy and cluster masses. \emph{Left:} The lensing [$M_{\mathrm{len}}(R \le 9.54\mathrm{\,kpc})$] and dynamical [$M_{\mathrm{dyn}}(R \le 9.54\mathrm{\,kpc})$] measurements of the galaxy projected total mass inside a 9.54 kpc radius as a function of the source redshift. The dashed and dotted lines represent 1 and 2 $\sigma$ error bars, respectively. \emph{Right:} The galaxy (dashed), the
cluster (dotted), and the total (dashed-dotted) projected mass inside
a 9.54 kpc radius centered on the galaxy. The long-dashed line shows
the projected mass of the cluster inside a circle with a radius of 488 kpc, that is
the distance between the cluster and the galaxy centers.}
\label{Mass_h}%
\end{figure}

A measurement of the value of the Einstein radius of a lensing system can be directly translated into a mass estimate for the deflector. In particular, the Einstein radius is defined as that radius inside which the mean surface density of a lens is equal to the critical surface density ($\Sigma_{\mathrm{cr}}$) of the system (for more details see Schneider et al. \cite{schneider}). So, it follows that the projected lensing mass ($M_{\mathrm{len}}$) enclosed within $R_{\mathrm{E}}$ equals
\begin{equation}
M_{\mathrm{len}}(\le R_{\mathrm{E}})=\Sigma_{\mathrm{cr}} \pi R^{2}_{\mathrm{E}}\,.
\label{eq:6}
\end{equation}
In our specific case, $\Sigma_{\mathrm{cr}}$ can assume a range of values depending on the source redshift. This is the main source of uncertainty in our lensing mass estimates, more relevant than that associated to detailed modeling of the cluster. From the results of Sect. 3.2 and 4, we have that the projected total mass of the lens galaxy without the cluster contribution $M_{\mathrm{len}}(R \le 9.54\mathrm{\,kpc})$ is well measured and its value is included between 4.6 and 6.2~$\times\mathrm{10^{11}\,M_{\sun}}$, at a 68\% CL. This information is shown in Table \ref{table:11} and on the left of Fig. \ref{Mass_h}.

\begin{table}
\caption{Luminous and dark mass measurements of the lens galaxy, without the cluster contribution. The ranges represent the intervals at 68\% confidence level.}
\label{table:11}
\centering
\begin{tabular}{c c c c c}
\hline\hline & $M_{\mathrm{len}}$ &  $M_{\mathrm{dyn}}$ & $M_{\mathrm{phot}}$ & $f_{\mathrm{DM}}$ \\ & $(\mathrm{10^{11}\,M_{\sun}})$ & $(\mathrm{10^{11}\,M_{\sun}})$ & $(\mathrm{10^{11}\,M_{\sun}})$ &  \\ \hline $R\leq8.10$ kpc & $3.9-5.2$ & $2.8-4.0$ & $0.6-0.9$ & $0.44-0.99$ \\ \hline $R\leq9.54$ kpc & $4.6-6.2$ & $3.3-4.7$ & $0.7-1.1$ & $0.50-0.99$ \\ \hline
\end{tabular}
\end{table}

The lower value of $R_{\mathrm{E_{G}}}$, found in Sect. 3.2 when considering the cluster mass contribution, suggests that the galaxy must be less massive than expected from the SIS+ES model. It is interesting to measure the projected total mass inside $R_{\mathrm{E}} = 9.54\mathrm{\,kpc}$, predicted by the SIS+ES and SIS+SIS models. In the former case, we use exactly Eq. (\ref{eq:6}); in the latter case, the mass is the sum of the cluster and of the
galaxy masses inside the same circle (see Fig. \ref{Mass_h}, on the right). The difference between the values of the total mass from the two models is less than
$0.3\%$. Therefore, the lensing mass estimate is found to be very
reliable and independent of the model details. To sum up, the galaxy
and the cluster masses inside the galaxy Einstein radius may have different weights, but their
sum is nearly constant. This explains the anti-correlation noted above
between the Einstein radius and the external shear, which is
interpreted as the cluster contribution. Moreover, the projected mass
of the cluster inside the circle of radius $R = 488\mathrm{\,kpc}$, the projected distance between the centers of the northern subcluster and the lens galaxy, agrees well with the value of about $3
\times \mathrm{10^{14}\,M_{\sun}}$ from analyses in X-rays (Huo et al. \cite{huo}), weak lensing (Jee et al. \cite{jee}), and dynamics of the cluster galaxies (Demarco et al. \cite{dem05}).

\subsection{Dynamical estimates}

Several studies (e.g., Rusin et al. \cite{rus03}; Koopmans et al. \cite{koo06}) have shown that the total (luminous+dark) density distribution of elliptical galaxies is homologous and well described by a $1/r^{2}$ (isothermal) profile. A similar conclusion was also reached by stellar dynamical studies (out to $R_{\mathrm{e}}$) of nearby galaxies, based on the application of self-consistent equilibrium dynamical models that incorporate the picture of galaxy formation by collisionless collapse (Bertin \& Stiavelli \cite{ber93}).

We consider here mass measurements projected along the line of sight. The projected mass ($M$) enclosed inside a certain radius ($R$) for an SIS is
\begin{equation}
M(\le R)=\frac{\pi\sigma_{\mathrm{SIS}}^{2}R}{G}\,,
\label{eq:5}
\end{equation}
where $\sigma_{\mathrm{SIS}}$ has been introduced in Eq. (\ref{eq:4}), and $G$ is the universal gravitational constant. As mentioned in Sect. 4.1, a good estimator of $\sigma_{\mathrm{SIS}}$ is found to be $\sigma_{0}$, and this latter quantity has been measured for the galaxy studied in this paper. Starting from Eq. (\ref{eq:5}) and the value of $\sigma_{0}$, we evaluate the dynamical projected total mass, $M_{\mathrm{dyn}}(\le R)$, within the two radii considered in the lensing analysis of the previous section. The results are presented in the third column of Table \ref{table:11} and on the left of Fig. \ref{Mass_h}.

The lensing and dynamical mass estimates are consistent and the lensing measurements do not turn out to be more accurate than the dynamical ones only because a rough estimate of the source redshift is available.

\subsection{Photometric estimates}

An estimate of the photometric-stellar mass ($M_{\mathrm{phot}}$) of a galaxy can be derived by comparing the observed SED of the galaxy with a set of composite stellar population (CSP) templates, computed with stellar population models. In addition to the stellar mass, this
method also allows the age and the star formation history (SFH) of the
galaxy to be investigated. It is well-known that the derived stellar mass depends on the adopted initial mass function (IMF) and only weakly on the assumed model of dust extinction and metallicity evolution. Moreover, it has been shown (Rettura et al. \cite{ret06}) that the photometric-stellar mass does not exhibit statistically significant discrepancies when evaluated with different stellar population models (e.g., Bruzual \& Charlot \cite{bru03} vs. Maraston \cite{mar05}). Finally, it must be noted that accurate stellar mass measurements require unbiased galaxy SEDs, which translates into accurate PSF-matched photometry.

\begin{figure}
\centering
\includegraphics[width=0.48\textwidth]{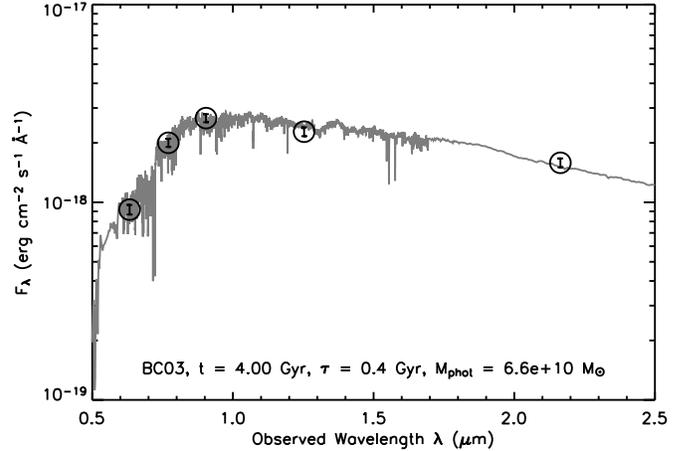}
\caption{SED of the lens galaxy at $z_{l}=0.82$. The circles with the error bars represent, from left to right, the observed flux densities measured inside the 1.07\arcsec aperture in the \emph{HST/ACS} (\emph{r}, \emph{i}, \emph{z}) and \emph{NTT/SofI} (\emph{J}, \emph{Ks}) passbands. The illustrated best-fit has been built with Bruzual \& Charlot \cite{bru03} models. On the bottom, the best-fit values of the age (t), the characteristic  time of the SFH ($\mathrm{\tau}$), and the mass (M$_{\mathrm{phot}}$) of the galaxy are given.}
\label{Mass_pho}%
\end{figure}

We derive the photometric-stellar mass of the lens galaxy through this multi-wavelength matched aperture photometry method. We use Bruzual \& Charlot's (\cite{bru03}) templates at solar metallicity, assuming a Salpeter (\cite{sal}) time-independent IMF and a delayed exponential SFH. First, we smooth all images (\emph{r}, \emph{i}, \emph{z} from \emph{HST/ACS} and \emph{J}, \emph{Ks} from \emph{NTT/SofI}) to the worst PSF and then we measure the galaxy magnitudes inside the apertures suggested by the lensing analysis (see Sect. 3.2). The results are summarized in Table \ref{table:10}. In Fig. \ref{Mass_pho} we plot the observed SED of the galaxy and the best-fit Bruzual \& Charlot \cite{bru03} model for the first aperture. The best-fit intervals, at the 68\% confidence level, of the stellar mass inside the two apertures are shown in the fourth column of Table \ref{table:11}.

\begin{table}
\caption{Multi-band photometry of the lens galaxy.}
\label{table:10}
\centering
\begin{tabular}{c c c c}
\hline\hline & & $\theta \leq 1.07$\arcsec & $\theta \leq 1.26$\arcsec \\ \hline \emph{HST/ACS} & $r_\mathrm{AB}$ & $23.688\pm0.014$ & $23.441\pm0.013$ \\ \hline \emph{HST/ACS} & $i_\mathrm{AB}$ & $22.415\pm0.005$ & $22.171\pm0.005$ \\ \hline \emph{HST/ACS} & $z_\mathrm{AB}$ & $21.755\pm0.004$ & $21.511\pm0.004$ \\ \hline \emph{NTT/SofI} & $J_\mathrm{AB}$ & $21.219\pm0.010$ & $20.990\pm0.009$ \\ \hline\emph{NTT/SofI} & $Ks_\mathrm{AB}$ & $20.426\pm0.011$ & $20.183\pm0.011$ \\ \hline
\end{tabular}
\end{table}

Starting from the lensing and dynamical measurements of the total mass and from the photometric measurement of the stellar mass, we have determined the minimum amount of dark matter ($f_{\mathrm{DM}}$). The radii relative to which $f_{\mathrm{DM}}$  is computed are about four times larger than the effective radius ($R_{\mathrm{e}}$) of the galaxy (see Table \ref{table:1}). In order to obtain the dark matter fraction, we assume the ``maximum light'' hypothesis: inside each aperture the stellar mass is divided by the lower total mass estimate (in this particular case, the latter value comes from the dynamical study). This gives an upper limit to the stellar mass fraction, from which a lower limit to the dark mass fraction can be inferred. In the last column of Table \ref{table:11} the measured values of $f_{\mathrm{DM}}$ are presented.

The total luminosity of the system inside the 9.54 kpc Einstein radius is $(3.1 \pm 0.3) \times 10^{10}\,L_{\odot,B}$, implying an average mass-to-light ratio $M/L_{B} = (13.0 \pm 2.6)\, M_{\odot}\,L_{\odot,B}^{-1}$ (considering the lower dynamical estimate of the total mass).
Under appropriate assumptions (e.g., see Treu et al. \cite{tre01, tre05}) the evolution of the intercept of the Fundamental Plane with redshift can be related to the evolution of the average effective mass-to-light ratio. The SLACS and LSD Surveys (Treu et al. \cite{tre06}) have established that, in the redshift range from 0 to 1, the effective stellar mass-to-light ratio evolves as $\mathrm{d} \log (M/L_{B})/\mathrm{d}z = -0.76$, with an rms scatter of 0.11.
If the evolution of the effective mass-to-light ratio is equal to the evolution of the stellar mass-to-light ratio, we can use this result to infer $M_{*}/L_{B}$ of our lens galaxy, assuming that $\log (M_{*}/L_{B})_{z} = \log (M_{*}/L_{B})_{0} + \Delta \log (M/L_{B})$.
The first term on the right-hand side of the previous equation can be measured for local E/S0 galaxies; e.g., using the data from Gerhard et al. \cite{ger} $(M_{*}/L_{B})_{0} = (7.9 \pm 2.3)\,M_{\odot}\,L_{\odot,B}^{-1}$.
Hence, we find that $M_{*}/L_{B}\,(z=0.82)= (1.9\pm0.7)\,M_{\odot}\,L_{\odot,B}^{-1}$. We note that the value $M_{*}/L_{B} = 13.0\,M_{\odot}\,L_{\odot,B}^{-1}$ required to explain the projected mass enclosed inside the Einstein radius solely by luminous matter is inconsistent with the above independently derived value at more than 4 $\sigma$ level. This is further evidence of the presence of a significant dark matter component inside the lens Einstein circle. Moreover, from the estimate of the photometric-stellar mass we obtain for $M_{*}/L_{B}$ a value between 2.3 and 3.5, which is consistent with the value derived from the evolution of the Fundamental Plane.

We mention that our findings on the fraction of the dark matter component are compatible with the results from other different studies on the amount of dark matter in elliptical galaxies (e.g., Saglia et al. \cite{sag92}; Treu \& Koopmans \cite{tre04}; Ferreras et al. \cite{fer05}; Treu et al. \cite{tre06}). We also note that our relatively high values of $f_{\mathrm{DM}}$ support the picture that the most massive ellipticals be dark matter dominated in their outer regions.

%

%

%

\section{Conclusions}
In this paper we have presented a strong lensing analysis of an
elliptical galaxy, member of the cluster Cl J0152.7-1357 ($z \simeq
0.84$). By means of \emph{HST/ACS} deep observations, a three-component
source highly magnified into twelve images was discovered. For this
system we have discussed several parametric macroscopic models: first point-like, then extended image models have been developed. 
The Einstein radius of a
singular isothermal sphere was found to be $R_{\mathrm{E}} = 9.54 \pm
0.15$ kpc, and the value of the projected mass inside this radius was shown to be a function of the unknown source redshift $z_{s}$.
By combining lensing with galaxy dynamics and by studying the source spectrum, we have obtained for this redshift lower and upper bounds of 1.9 and 2.9, respectively. These limits have allowed us to get accurate and robust total mass measurements of the galaxy: $M_{\mathrm{len}}(R \le
9.54\mathrm{\,kpc})=(4.6-6.2)\,\times\mathrm{10^{11}\,M_{\sun}}$.
Furthermore an external shear component was proved to be indicative of
the northern mass distribution of the cluster and the predicted value
of the cluster mass has been shown to be in agreement with the X-ray, weak lensing, and cluster member dynamical analyses.
Then, we have measured the lens total and luminous mass from stellar dynamics, and from optical and near-IR photometry. From these results we have estimated a lower limit of 50\% (1 $\sigma$) for the dark matter fraction enclosed inside the Einstein radius of the galaxy. The presence of a significant dark matter component is also confirmed by comparing the value of the mass-to-light ratio measured in our lens with the stellar one predicted from the evolution of the Fundamental Plane.

Using non-parametric methods (see Brada{\v c} et al. \cite{brad}), the
complex image configuration of the lensing system might be better
reproduced, but possibly with only a small gain of insight into the lens properties.

\begin{acknowledgements}
      Part of this work was supported by the \emph{European Southern
      Observatory Director General Discretionary Fund}.
\end{acknowledgements}

\end{document}